\documentclass[aps,prl,twocolumn,longbibliography,showpacs,amsmath,amsmath,amssymb,amsfonts,superscriptaddress]{revtex4-1}
\usepackage{graphicx}
\usepackage{dcolumn}
\usepackage{bm}
\usepackage{color}
\usepackage{multirow}
\usepackage{natbib}
\usepackage{hyperref}
\usepackage{dsfont}
\usepackage{bbold}

\newcommand{\lan}{\left\langle}
\newcommand{\ran}{\right\rangle}

\newcommand{\sbigotimes}{%
  \mathop{\mathchoice{\textstyle\bigotimes}{\bigotimes}{\bigotimes}{\bigotimes}}%
}

\newcommand{\mal}{\mathcal}

\newcommand{\ha}{\frac{1}{2}}
\newcommand{\la}{\langle}
\newcommand{\ra}{\rangle}
\newcommand{\be}{\begin{equation}}
\newcommand{\ee}{\end{equation}}

\newcommand{\ket}[1]{\left| #1 \ran}
\newcommand{\bra}[1]{\lan #1 \right|}

\newcommand{\comm}[2]{\left[ #1, #2 \right]}
\newcommand{\acomm}[2]{\left\{ #1, #2 \right\}}

\begin{document}

\title{Glassy dynamics due to a trajectory phase transition in dissipative Rydberg gases}

\author{Carlos P\'erez-Espigares}
\affiliation{School of Physics and Astronomy and Centre for the Mathematics and Theoretical Physics of Quantum Non-equilibrium Systems,
University of Nottingham, Nottingham NG7 2RD, UK}
\author{Igor Lesanovsky}
\affiliation{School of Physics and Astronomy and Centre for the Mathematics and Theoretical Physics of Quantum Non-equilibrium Systems,
University of Nottingham, Nottingham NG7 2RD, UK}
\author{Juan P. Garrahan}
\affiliation{School of Physics and Astronomy and Centre for the Mathematics and Theoretical Physics of Quantum Non-equilibrium Systems,
University of Nottingham, Nottingham NG7 2RD, UK}
\author{Ricardo Guti\'errez}
\affiliation{School of Physics and Astronomy and Centre for the Mathematics and Theoretical Physics of Quantum Non-equilibrium Systems,
University of Nottingham, Nottingham NG7 2RD, UK}
\affiliation{Complex Systems Group \& GISC, Universidad Rey Juan Carlos, 28933 M\'ostoles, Madrid, Spain.}


\date{\today}
\begin{abstract}
The physics of highly excited Rydberg atoms is governed by blockade or exclusion interactions that hinder the excitation of atoms in the proximity of a previously excited one. This leads to cooperative effects and a relaxation dynamics displaying space-time heterogeneity similar to what is observed in the relaxation of glass-forming systems. Here we establish theoretically the existence of a glassy dynamical regime in an open Rydberg gas, associated with phase coexistence at a first-order transition in dynamical large deviation functions. This transition occurs between an active phase of low density in which dynamical processes take place on short timescales, and an inactive phase in which excited atoms are dense and the dynamics is highly arrested. We perform a numerically exact study and develop a mean-field approach that allows us to understand the mechanics of this phase transition. We show that radiative decay --- which becomes experimentally relevant for long times --- moves the system away from dynamical phase coexistence. Nevertheless, the dynamical phase transition persists and causes strong fluctuations in the observed dynamics.
\end{abstract}


\maketitle

{\it Introduction} -- The study of cold-atomic ensemble dynamics allows the exploration of a vast array of many-body effects relevant to condensed-matter and statistical physics in settings accessible to modern experiments. Among these systems, cold gases of highly excited (Rydberg) atoms constitute a versatile platform due to their strong and tunable interactions \cite{gallagher2005,low2012}. The competition between excitation and interaction in these systems leads to complex collective behavior, which is currently intensely explored both experimentally \cite{urvoy2015,valado2016,gutierrez2017} and theoretically \cite{lesanovsky2013, lesanovsky2014,mattioli2015classical, angelone2016superglass,perez-espigares2017,letscher2017bistability}. 
Particularly interesting is that the physics of these systems is governed by blockade effects reminiscent of the excluded volume effects of classical many-body systems close to the glass and jamming transitions \cite{binder2011,sanders2015sub} and their idealization as kinetic constraints \cite{ritort2003}. The resulting dynamics is highly heterogeneous, with regions that evolve very rapidly while others remain stuck in their local configurations for long times \cite{lesanovsky2013,gutierrez2015}. 

\begin{figure}[h!]
\vspace{-0.35cm}\includegraphics[scale=0.475]{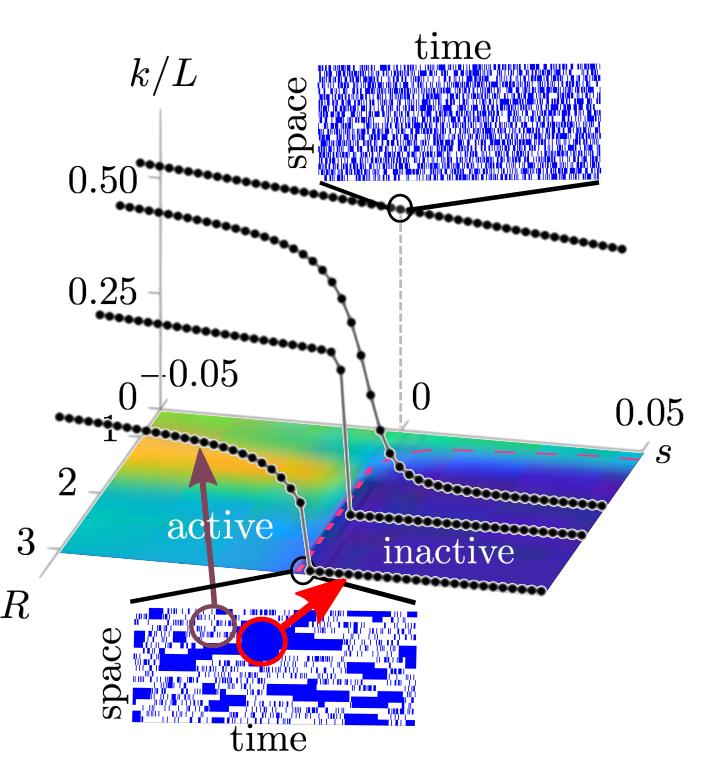}
\vspace{-0.4cm}
\caption{{\sf \bf Dynamical first-order phase transition underlying the dynamics of dissipative Rydberg gases.}
Activity $k(s)/L$ in a system of $L=15$ atoms as a function of the tilting field $s$ and the blockade length $R$. Black curves correspond to $R=1, 1.5, 2$ and $3$, while results for a range of $R$ values are displayed in the 2D color map at the base (the red dashed line shows the position of the inflection points).  Representative trajectories for $R=1$ (upper panel) and $R=3$ (lower panel) for $L=20$ are displayed. Blue and white indicate excited and ground state atoms, respectively.}\label{fig1}
\end{figure}

In this work, we establish on firm grounds that the heterogeneous dynamical behavior of Rydberg gases is due to the emergence of a glassy regime which stems from an underlying dynamical phase transition. The latter, associated with sudden structural changes in the trajectories of the system, is analyzed following a ``thermodynamics of trajectories'' approach \cite{ruelle2004thermodynamic,merolle2005,garrahan2007,lecomte2007,garrahan2018}, which is based on a large-deviation principle \cite{eckmann1985,dembo1998,touchette2009}. This approach unveils the structure and dynamics adopted by physical systems in order to sustain unlikely values of certain observables over long periods of time. Here we shall focus on the activity \cite{garrahan2007, lecomte2007,baiesi2009}, which counts the number of atoms that (de-)excite per unit time, and is a relevant observable in the context of cold-atomic ensembles, as it can be experimentally accessed by the continuous observation of light scattered off the atoms \cite{mekhov2007cavity}. The dynamical heterogeneity of Rydberg gases, observed in the glassy regime, is explained by the fact that the system lies at a coexistence point between an active and an inactive phase, see Fig.~\ref{fig1} (to be discussed below).

The inactive space-time regions that appear as the transition is approached from the active side, are ``bubbles of inactivity'',  corresponding to a manifestation in trajectories of fluctuations associated with the dynamical first-order transition (cf. e.g. vapor bubbles in a liquid near liquid-vapor coexistence). These dynamical fluctuations are similar to those observed in more traditional glass forming systems \cite{chandler2010,katira2017}.

Generally speaking, the study of dynamical fluctuations in terms of large deviations serves two purposes. On the one hand, it shows how to engineer the dynamics of a system so that trajectories display desirable features ---such as ordered sequences or bursts of photons--- associated with rare events taking place far away from the typical dynamics \cite{Garrahan2010,jack2010a,Touchette2013a,Chetrite2015,JackSollich2015,carollo2017b}. In fact, the system can be conditioned to have specific output statistics not only in the asymptotic time limit but also for transient dynamics, which may be more easily accessible experimentally than stationary states \cite{carollo2017b}. On the other hand, the large deviation approach allows us to explain intriguing phenomena such as intermittent dynamical regimes \cite{garrahan2007,ates2012b}, or the glassy regime of dissipative Rydberg gases that is studied in this paper.

Our goal is to establish the existence of the phase transition that underlies the behavior of dissipative Rydberg gases, and to discuss its main features. To this end, we perform a detailed investigation of the dynamical phase diagram as a function of the interaction strength, as well as in the presence of radiative decay processes. The latter are particularly important in experiments monitoring long-time behavior. Beyond numerical simulations we develop a mean-field approach which leads to an understanding of the phase transition mechanism in terms of a dynamical free energy. Our work consolidates the understanding of collective dynamical phenomena --- such as dynamical heterogeneity --- in dissipative Rydberg gases and establishes a direct connection to soft-matter physics and the physics of glassy matter in particular. We expect our analysis to be applicable for uncovering collective dynamical behavior also in other driven dissipative spin systems, such as e.g. systems of interacting nitrogen-vacancy centres \cite{zhang2017,jelezko2006} or electrons and nuclei in non-equilibrium nuclear magnetic resonance \cite{abragam1978}.

{\it Model} -- We consider a driven dissipative system of highly-excited (Rydberg) atoms in a 1D chain. At each of the $L$ sites lies an atom that can be in its ground state $\ket{\downarrow}$ or in a high-lying (Rydberg) excited state $\ket{\uparrow}$. The interaction potential  between atoms $j$ and $k$ is non-negligible only if both are in the excited state, and is given by $V_{jk} = C_\alpha/|r_j - r_k|^\alpha$, where $r_j$ gives the position of $j$ in units of the lattice spacing. The $\ket{\downarrow} \leftrightarrow \ket{\uparrow}$ transition is resonantly driven by a laser field with Rabi frequency $\Omega$. The coherent part of the dynamics is thus generated by the Hamiltonian $H = \frac{1}{2}\sum_{j=1}^L \sum_{k=1}^L V_{jk} n_j n_k + \Omega \sum_{j=1}^L \sigma_j^x$, where $n_j = \ket{\uparrow}_j \bra{\uparrow}$ and $\sigma_j^x = \ket{\uparrow}_j \bra{\downarrow} +\ket{\downarrow}_j \bra{\uparrow}$. Further, we consider that due to different forms of noise (laser linewidth, thermal effects, etc.~\cite{low2012,urvoy2015,valado2016,gutierrez2017}) the system is affected by dephasing with a rate $\gamma$. The evolution of the system is then governed by the Lindblad equation $\partial_t \rho = - i\comm{H}{\rho} + \sum_{j=1}^L \mal{L}(\sqrt{\gamma}\, n_j) \rho$. Here, $\mal{L}(J) \rho = J \rho J^\dag - \ha \acomm{J^\dag J}{\rho}$ is a dissipator in Linblad form with jump operator $J$. In the large dephasing limit, $\gamma \gg \Omega$, there is a clear separation between the timescales on which quantum coherences are produced and destroyed, which allows to perform an adiabatic elimination of quantum coherences \cite{ates2007,lesanovsky2013,cai2013,marcuzzi2014,honing2014,vsibalic2016driven}. The resulting dynamics is governed by a classical master equation with configuration-dependent rates for transitions $\ket{\downarrow}_j\leftrightarrow\ket{\uparrow}_j$
\begin{equation}
\Gamma_j=  \frac{4 \Omega^2}{\gamma}\frac{1}{1+\left(R^{\alpha} \sum_{k\neq j} \frac{n_k}{|r_k - r_j|^\alpha} \right)^2}, 
\label{rate}
\end{equation}
with the interaction parameter $R = (2 C_\alpha/\gamma)^{1/\alpha}$ giving the length of the blockade radius \cite{lesanovsky2013,gutierrez2015}. The validity of this approach has been confirmed in recent experiments \cite{schempp2014,urvoy2015,valado2016}.  Below, we consider the experimentally relevant case of van der Waals interactions, $\alpha = 6$, and rescale the rates ${\tilde \Gamma}_j=\gamma \Gamma_j/(4\Omega^2)$ so that the time unit reflects the excitation timescale in the absence of interactions.

{\it Dynamical phase transition: numerically exact study of finite-size systems} --  
We start with a numerical exploration of the stationary dynamics of the system.
In the upper and lower part of Fig.~\ref{fig1}, we show representative trajectories for $R=1$ and $R=3$ respectively obtained via continuous-time Monte Carlo simulations \cite{bortz1975,newman1999}. While for $R=1$ the dynamics appears to be homogeneous and characterized by a single time scale, for larger values of $R$ we observe an alternation of relatively dilute regions that evolve quite rapidly and regions formed by blocks of contiguous excitations that evolve along much longer time scales. The latter observation is confirmed by the visual inspection of trajectories and two-time correlation functions $\langle n_j(t) n_j(0)\rangle$, which display prominent plateaus indicative of the existence of more than one timescale \cite{binder2011}.

This phenomenology is shown to arise from a dynamical trajectory phase transition. This kind of transitions have been previously studied in a large family of models of glassy soft matter \cite{garrahan2007, garrahan2009}, but had never before been previously identified in dissipative Rydberg gases. Such transitions are probed through a dynamical observable ---the activity per unit time $k=K/t$ \cite{garrahan2007, lecomte2007,baiesi2009}, which quantifies the number of state changes (spin flips) $K$ in a trajectory of duration $t$. The phase diagram corresponding to this dynamical order parameter is obtained via a thermodynamics of trajectories approach \cite{ruelle2004thermodynamic,eckmann1985}, which allows us to infer the presence of dynamical phase transitions through the structure of the probability distribution of the activity $P(k)$. For long times, this probability satisfies a large deviation principle $P(k)\approx e^{-t \phi(k)}$, where $\phi(k)$ is the so-called large deviation function
\cite{garrahan2009}.
Often it is more convenient to study the moment generating function $Z(s)=\la e^{-s t k}\ra\approx e^{t \theta(s)}$, where the scaled cumulant generating function (SCGF) $\theta(s)$ is related to $\phi(k)$ via a Legendre transformation  \cite{touchette2009}. Here $Z(s)$ is a dynamical partition function of an ensemble of trajectories ---the $s$-ensemble (see \cite{garrahan2009} and the recent review \cite{garrahan2018}). In this framework, $t$ plays the role of the volume in equilibrium and $k$ is an order parameter ---analog of the internal energy density--- with a conjugate field $s$ acting as a control parameter. Notice that this parameter (which is not experimentally accessible) allows for the exploration of dynamical regimes that are unlikely to be observed as spontaneous fluctuations, since their probability decays exponentially in time. Such regimes can be attained, however, by suitably engineering the dynamics of the system \cite{Garrahan2010,jack2010a,Touchette2013a,Chetrite2015,JackSollich2015,carollo2017b}.
 
We will focus our study on the SCGF $\theta(s)$,
whose non-analyticities correspond to phase transitions between dynamical phases. 
It is 
given by the eigenvalue with the largest real part of the so-called tilted generator \cite{touchette2009, garrahan2018}
\be
{\mathbb W}_s=\sum_{j=1}^L \left[{\tilde \Gamma}_j(e^{-s} \sigma_j^x - 1)  \right]\, ,
\label{tiltop}
\ee
that governs the evolution biased by the field $s$ \cite{garrahan2009}, which ``tilts'' the systems towards more (if $s<0$) or less (if $s>0$) active dynamics. For $s=0$, the unbiased dynamics of the (unperturbed) system is recovered.

In Fig.~\ref{fig1} we show the activity density $k(s)/L$ for a range of values of $R$. We observe a qualitative change as $R$ is varied at $s=0$: whereas for small $R$ the change from active ($s<0$) to inactive ($s>0$) is smooth, a discontinuity develops in the system for $R\geq 2$. The jump becomes smaller as $R$ is increased, as the activity for $s<0$ decreases (i.e., when the blockade radius is larger, the overall activity is lower). This discontinuity indicates the presence of a first-order dynamical phase transition \cite{garrahan2009}, which is below theoretically established at the mean-field level. The dynamics observed in the trajectory for $R=3$ (and for other values of $R>2$ that we have numerically explored) arises from the coexistence of space-time regions, with low-activity regions playing the role of the familiar bubbles of an equilibrium transition in a fluid.

\begin{figure}[t!]
\hspace{-0.285cm}\includegraphics[scale=0.110]{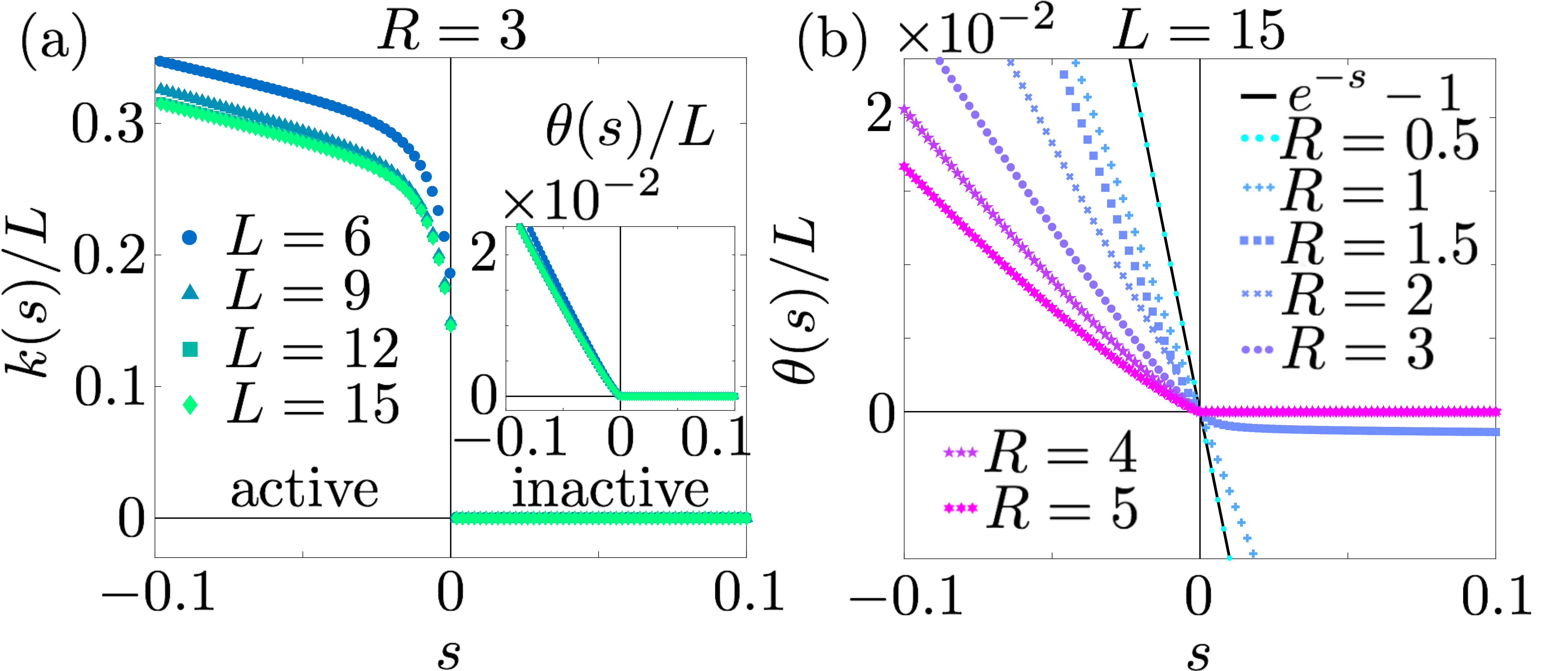}
\caption{{\sf \bf Scaled cumulant generating function and activity for finite-size systems.}
(a) Activity density $k(s)/L$ (main panel) and scaled cumulant generating function $\theta(s)/L$ (inset) as functions of the tilting field $s$ for $R=3$ and system sizes $L=6$, $9$, $12$ and $15$. (b) Scaled cumulant generating function $\theta(s)$ in a system of $L=15$ sites for $R=0.5$, $R=1$, $R=1.5$, $2$, $3$, $4$ and $5$. For $R=0.5$, the curve closely follows the Poissonian form $e^{-s}-1$, see black line.
}
\label{fig2}
\end{figure}

The activity density $k(s)/L$ shown in Fig.~\ref{fig1} is obtained from the SCGF $\theta(s)$ through the relation $k(s)/L = - \theta^\prime(s)/L$ \cite{garrahan2009}. The numerical diagonalization of the tilted generator ${\mathbb W}_s$ needed for obtaining the SCGF has been performed for a system of size $L=15$, as larger sizes are numerically prohibitive due to the exponential growth of the generator. We illustrate the finite-size behavior of the activity density for $R=3$ in Fig.~\ref{fig2} (a), where $k(s)/L$ is seen to converge to a size-independent curve for $L>10$. A similar dependence on $L$ is observed for the whole range of the interaction parameter $R$. The SCGF itself is shown in Fig.~\ref{fig2} (b), also for $L=15$, which displays a smooth dependence on $s$ for $R<2$, and becomes Poissonian for $R=0.5$, when the system is virtually non-interacting.

\begin{figure*}[t!]
\hspace{-0.2cm}\includegraphics[scale=0.107]{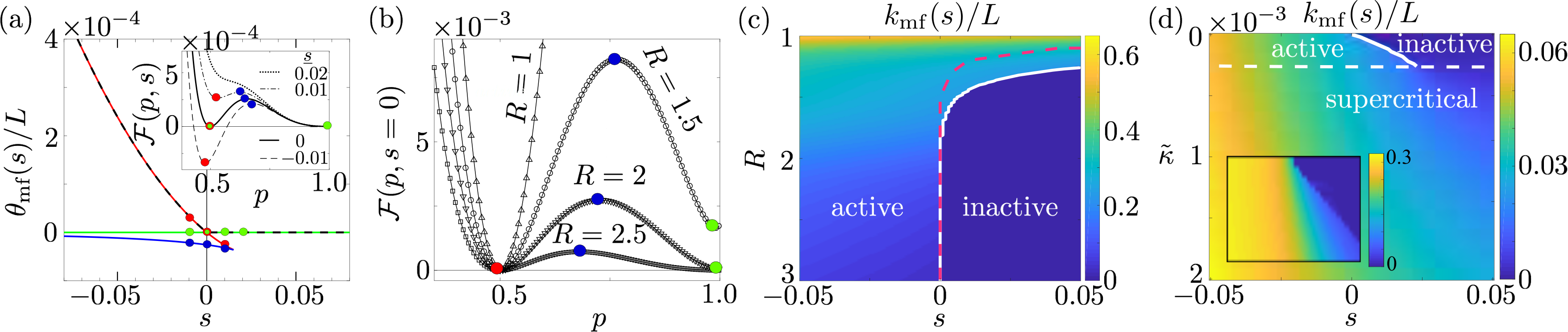}\\
\caption{{\sf \bf Mean-field analysis of the dynamical phase transition.}
(a) (Negative) variational free energy $-{\cal F}(p,s)$ for $R=3$ evaluated at the stationary points including two maxima (red and green lines) and one minimum (blue line), and (normalized) SCGF $\theta_\text{mf} (s)/L$ (dashed black line). {\it Inset}: Variational free energy ${\cal F}(p,s)$ as a function of $p$ in the neighborhood of  $s=0$. (Red, blue, and green discs highlight the correspondence between points in the main panel and the inset.) (b) Variational free energy ${\cal F}(p,s=0)$ for values of $R$ around the critical value for a transition at $s=0$. (c) Activity $k_{\text{mf}}(s)/L$ as function of the tilting field $s$ and $R$. (d) Activity $k_{\text{mf}}(s)/L$ as a function of the (rescaled) decay rate $\tilde\kappa$ and $s$ for $R=3$. The solid white line indicates the position of the phase transition where both phases coexist, which ends at a critical point $\tilde \kappa_c \approx 2.8 \times 10^{-4}$. Beyond this point, a smooth crossover is observed. {\it Inset}: Activity density of the full dynamics $k(s)/L$ in a system of $L=12$ atoms for the same range of $s$ and $\tilde\kappa$. 
}\label{fig3}
\end{figure*}

{\it Dynamical phase transition: mean-field analysis} -- To shed further light on this dynamical behavior, we introduce a variational free energy defined as ${\cal F}_{\ket{V}}(s) = -\langle V|{\mathbb W}_s|V\rangle$, for normalized states $\ket{V}$, whose minima correspond to dynamical phases \cite{garrahan2009}. In our mean-field approach, we focus on the subspace of (uncorrelated) states $\ket{V}= \bigotimes_{j} \ket{v}_j$ for $\ket{v}_j = \sqrt{p} \ket{\uparrow}_j + \sqrt{1-p} \ket{\downarrow}_j$. Since states are parametrized by the density of excitations $p$, the free energy will be denoted as ${\cal F}(p,s)$.  ${\mathbb W}_s$ is symmetric, and the SCGF is approximated by minus the minimum of the variational free energy, $\theta_\text{mf} (s)= \max_{\ket{V}} [\langle V|{\mathbb W}_s|V\rangle]=-\min_{p} [{\cal F}(p,s)]$, with
\begin{eqnarray}
{\cal F}(p,s)\! &=&\! -\sum_{j=1}^L(\sbigotimes_{l\neq j} {}_l\!\bra{v}) {\tilde \Gamma}_j  (\sbigotimes_{m \neq j}\! \ket{v}_m) {}_j\!\bra{v}\! (e^{-s}\sigma_j^x - 1)\!\ket{v}_j \nonumber\\
&=& -L\, \Gamma_\text{mf}\!\left[2  \, e^{-s}\! \sqrt{p (1\!-\!p)}  - 1\right]\!.
\label{sandwich}
\end{eqnarray}
This expression depends on the (de-)excitation rates
\begin{equation}
\Gamma_\text{mf} =\! \sum_{c_1=0}^2 \sum_{c_2=0}^2 \cdots\! \sum_{c_{r }=0}^2 \frac{\displaystyle 2^{\sum_{l=1}^{r }\! \delta_{c_l, 1}} p^{\sum_{l=1}^{r }\! c_l} (1-p)^{2 r  - \sum_{l=1}^{r }\! c_l}}{1 + R^{2\alpha} \left(\sum_{l=1}^{r } \frac{c_l}{l^\alpha}\right)^2},
\end{equation}
where $l=1, 2, \ldots, r $ are the distances to the first, second,..., $r$-th nearest neighbors. The coefficients $c_l$ denote occupation numbers, i.e., $c_l = 0$ if neither of the $l$-th nearest neighbors are excited, $c_l  = 1$ if one of them is, and $c_l = 2$ if both are. Strictly speaking, one should consider the limit $r \to \infty$ in the various summations, but in practice truncating the sum to a value somewhat larger than $R$ should suffice, given the $1/l^\alpha$ decay. (In what follows, we set $r  = 7$  for $\alpha=6$ and $R\leq 3$, as the results do not change apprecibly for larger $r$.) The factor $2^{\sum_{l=1}^{r }\! \delta_{c_l, 1}}$ ($\delta_{c_l, 1} = 1$ if $c_l = 1$, and is zero otherwise) accounts for the degeneracy of the $c_l = 1$ case ($c_l=1$ if the left neighbor is excited, but also if the right one is).

The extrema of ${\cal F}(p,s)$ with respect to the excitation density $p$ are found numerically. For $R\lesssim 1.25$ there is only one (real) solution, corresponding to a minimum, while for $R>1.25$ there are two minima and one maximum. 
In the main panel of Fig.~\ref{fig3} (a)  we show $-{\cal F}(p,s)$ for $R=3$ evaluated at each of these extrema  as a function of $s$ (continuous lines), and the SCGF $\theta_\text{mf} (s)$ (black dashed line). In the inset, the smaller minimum of ${\cal F}(p,s)$ for $s<0$ is seen to become larger than the other one for $s>0$, which results in the non-analytic behavior of $\theta_\text{mf} (s)$ (i.e., the absolute minimum) at $s=0$, where the two minima are equally deep. Furthermore, we see a discontinuity in the density of excitations, which goes from $p\approx 0.5$ for $s<0$ to $p\approx 1$ for $s>0$, with a metastable region for $s\gtrsim 0$ that ends in a spinodal point where the local minimum merges with the maximum. This establishes the existence of a first-order dynamical phase transition at $s=0$ which was suggested by the numerical results of Fig.~\ref{fig1}.

{\it Phase diagram} -- We next focus on the dependence of the free energy on the blockade length $R$ which is shown in Fig.~\ref{fig3} (b). This data shows that the glassy phase is entered beyond $R=2$. For larger values of $R$ two minima are present  at $s=0$, indicating the coexistence of active and inactive regions, while for smaller values of $R$ there is just a single global minimum. In the latter regime the dynamics is largely uncorrelated (see Fig.~\ref{fig1}, upper panel), and the statistics of the activity become effectively Poissonian for $R<1$. 
For $1.25\lesssim R\lesssim 2$ a second local minimum develops that becomes as deep as the first one for slightly positive $s$. As a result, the transition appears slightly away from the unbiased dynamics at $s=0$, signaling the presence of fat tails in the probability distribution of the activity due to the strong fluctuations encountered for positive values of $s$. For larger $R$ the transition moves towards $s=0$, and saturates at $R\approx 2$, as shown by the white solid line of Fig.~\ref{fig3} (c), where the activity $k_\text{mf}(s)/L$ as a function of $s$ for a range of $R$ is displayed. The inflection points of the finite-size $k(s)/L$ curves of Fig.~\ref{fig1} are included for comparison (see the red dashed line), showing a good qualitative agreement with the mean-field results.

In the experimental study of Rydberg gases one inevitably faces radiative decay ($\ket{\uparrow} \rightarrow \ket{\downarrow}$). Theoretically, decay is accounted for by a set of jump operators $J=\sqrt{\kappa}\, \sigma^{-}_j$, where $\sigma_j^- = \ket{\downarrow}_j \bra{\uparrow}$, acting on each site $j=1,...,L$, for a given atomic decay rate $\kappa$ \cite{breuer2002}.
Fig.~\ref{fig3} (d) shows that a non-zero (rescaled) decay rate $\tilde \kappa = \gamma \kappa/(4\Omega^2)$ moves the phase transition line slightly from $s=0$ towards the $s>0$ region, indicating again strong dynamical fluctuations in such region both in the mean-field dynamics (main panel) and in the full dynamics for $L=12$ (inset).
Note that the variational mean-field analysis requires the inclusion of the process $\ket{\downarrow} \rightarrow \ket{\uparrow}$ along with decay in order to preserve the symmetry of the dynamical generator (see \cite{garrahan2009} for a detailed discussion). Both the exact numerics and the mean-field analysis show that the dynamical phase transition between regions of high and low activity $k$ is smoothed out beyond a critical decay rate, corresponding to $\tilde \kappa_c \approx 2.8 \times 10^{-4}$ in the mean-field dynamics and to a larger value in the full dynamics. For a discussion of dynamical critical points and the observability of inactive dynamics for phase transitions occuring at $s>0$, see Ref.~\cite{elmatad2010}. In an experiment, one would have to apply a sufficiently large Rabi frequency so that $\tilde \kappa$ remains below the critical value in order to clearly see the effect of the transition. For example, a moderate increase of the Rabi frequency of less than one order of magntiude would bring the system of rubidium atoms of Ref. \cite{gutierrez2017} into the desired regime.
If the decay rate exceeds the critical value, however, the transition between high activity and low activity becomes a smooth crossover, and the observability of inactive space-time regions in the trajectory depends in a non-trivial manner on how close one is to the critical point (as this is expected to influence the width of the free energy minimum, which determines the size of the fluctuations).

{\it Characterization of the dynamical phases} -- 
We now set out to describe the nature of the phases involved in the transition more quantitatively. To this end, we focus on the dynamics of finite-size systems without decay ($\tilde\kappa = 0$).
We first study the time average of the density of excitations $n = \frac{1}{t} \int_0^t d\tau \left( \frac{1}{L} \sum_{j=1}^L  n_j (\tau) \right)$ as a function of the tilting field $s$. Here the density of site $j$ is given by $n_j (\tau) = \langle - | n_j |C(\tau)\rangle$, where $C(\tau)$ is the configuration at time $\tau$ and $|- \rangle = \sum_{C^\prime} |C^\prime\rangle$. The average of the time-integrated observable
is equivalent to a static average
$\la n \ra_s = \la V_s|\frac{1}{L} \sum_{j=1}^L  n_j| V_s\ra,$
where $|V_s\ra$ is the eigenvector associated with the largest eigenvalue $\theta(s)$ of the tilted generator in Eq. \eqref{tiltop} \cite{garrahan2009}. This static average
can be generalized to two-point (or multi-point) observables, such as $\la n_i n_j \ra_s $. 
For the unbiased dynamics, $\ket{V_{s=0}} =\bigotimes_{j} 1/\sqrt{2}\, (\ket{\uparrow}_j +  \ket{\downarrow}_j)$, and thus $\langle n\rangle_{s=0} = 1/2$, for any value of $R$, as corresponds to a completely mixed stationary state, $\rho_\text{st} = 2^{-L} \bigotimes_j \mathds{1}_j$. \cite{lesanovsky2013}. 

\begin{figure}[t!]
\hspace{-0.25cm}\includegraphics[scale=0.112]{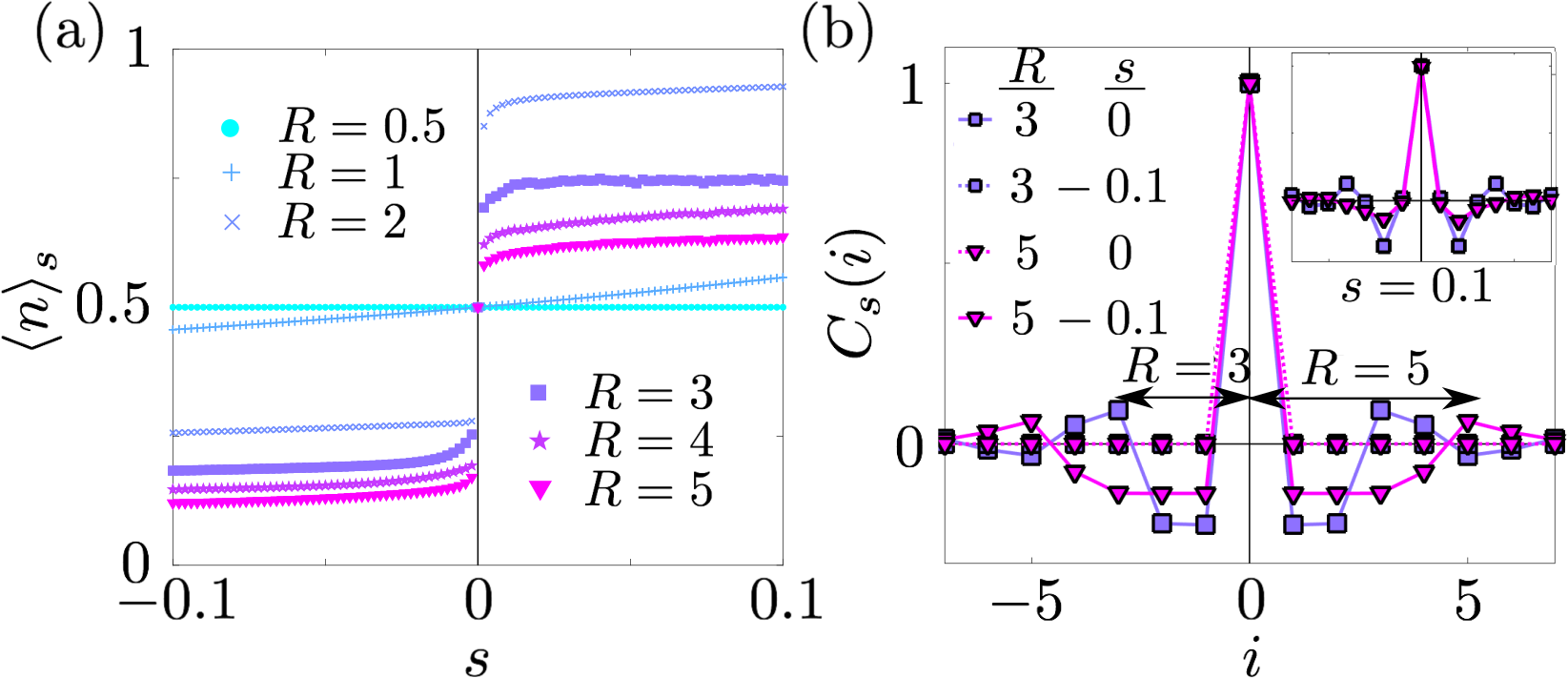}
\caption{{\sf \bf Density of excitations and spatial correlations for a system of $L=15$ atoms.}
(a) Average density of excitations $\langle n \rangle_s$ as a function of $s$ for $R= 0.5$, $1$, $2$, $3$, $4$ and $5$. (b) Normalized correlations $C_s(i)$ in the active phase ($s=-0.1$), for the unbiased dynamics ($s=0$) and in the inactive phase ($s=0.1$, see inset), for $R=3$ and $R=5$.
}
\label{fig4}
\end{figure}

The average density of excitations $\langle n\rangle_s$ for a range of values of $R$ shown in Fig.~\ref{fig4} (a) gives us the structural counterpart of the dynamical picture in Figs.~\ref{fig1} and \ref{fig3} (c). A high density of excitations corresponds to a low-activity dynamics, and viceversa.
In the glassy regime, for $R\gtrsim2$, the phase transition at $s=0$ separates an active phase with a very low density of excitations for $s<0$ from an inactive phase with a high density of excitations for $s>0$, with the overall density decreasing for larger blockade radius $R$. Beyond the critical value, for $R=1$ the dependence of the density with $s$ is smooth and mild, while for even smaller $R$ (i.e., when the atoms are essentially uncoupled) $\langle n\rangle_s = 1/2$ holds far from $s=0$.

To gain further insight into the structure of the two phases that coexists in the glassy regime, we next consider the (normalized) spatial correlations of the excitation density
$C_s(i) = (\langle n_i n_0 \rangle_s - \langle n \rangle_s^2)/(\langle  n \rangle_s - \langle n \rangle_s^2)$,
where $n_0$ is the number operator of the site taken as reference (due to translational invariance it can be any). These correlations, which characterize the spatial distribution of excitations, are shown in Fig.~\ref{fig4} (b) for a system of size $L=15$, with $R=3$ and $5$ and tilting field values around $s=0$ . For the the unbiased dynamics ($s=0$) the correlations are trivially given by the Kronecker delta $C(i) = \delta_{i,0}$ \cite{lesanovsky2013}. In the active phase ($s=-0.1$), however, there is a clear connection between the size of the anti-correlated region and the blockade length $R$: excitations tend to appear beyond the blockade radius, so as to maximize the activity \cite{gutierrez2015}. In the inactive phase (see inset for $s=0.1$), the overall activity is reduced by the disappearance of anti-correlations between nearest neighbors. This makes it possible to find contiguous blocks of excitations in the dynamics, as those shown in the inactive space-time regions of the $R=3$ trajectory in Fig.~\ref{fig1} (lower panel).

{\it Outlook} -- We have established the existence of a dynamical first-order phase transition at the level of trajectories in dissipative Rydberg gases, in analogy with classical glass formers \cite{garrahan2007,hedges2009,speck2012}.  In this framework, the arrested space-time regions observed in a background of more highly active dynamics in the simulated trajectories are the natural manifestation of a coexistence between an inactive and an active phase. The transition is shown to persist in the presence of atomic decay, and should therefore be observable in cold atoms experiments that continuously monitor the evolution of the atomic density e.g. via light scattering \cite{mekhov2007light}.

\begin{acknowledgments}
We thank Federico Carollo and Matteo Marcuzzi for insightful discussions. The research leading to these results has received funding from the European Research Council under the European Union's Seventh Framework Programme (FP/2007-2013) / ERC Grant Agreement No. 335266 (ESCQUMA) and the EPSRC Grant No. EP/M014266/1. R.G. acknowledges the funding received from the European Union's Horizon 2020 research and innovation programme under the Marie Sklodowska-Curie grant agreement No. 703683. I.L. gratefully acknowledges funding through the Royal Society Wolfson Research Merit Award. We are also grateful for access to the University of Nottingham High Performance Computing Facility, and for the computational resources and assistance provided by CRESCO, the super-computing center of ENEA in Portici, Italy.
\end{acknowledgments}

\bibliography{LDRydberg}{}

\end{document}